# Metastable polaron-supporting phase in poly p-phenylene vinylene films induced by UV illumination

T. Drori[1,2,4], E. Gershman[1,3], C.X. Sheng[4], Y. Eichen[1,3], Z.V. Vardeny [2,4], E. Ehrenfreund[1,2]

[1]Solid State Institute, Technion–Israel Institute of Technology, Haifa 32000, Israel
[2]Department of Physics, Technion–Israel Institute of Technology, Haifa 32000, Israel
[3]Department of Chemistry, Technion–Israel Institute of Technology, Haifa 32000, Israel
[4]Department of Physics, University of Utah, Salt Lake City, UT 84112, USA

We discovered a new metastable polaron-supporting phase in pristine films of a soluble derivative of poly-p-phenylene vinylene (MEH-PPV) that is induced by UV illumination. In the initial un-illuminated phase *A*, the films do not show long-lived photogenerated polarons. However, prolonged UV illumination for several hours induces a reversible, metastable phase *B* that shows abundant long-lived photogenerated polarons. Phase *B* films transform back to the original phase A within ½ hour in the dark at room temperature. We propose a reversible mechanism in which UV illumination creates metastable deep defects that substantially increase the photogenerated polaron lifetime.

Photomodulation (PM) spectroscopy has served as a useful tool to investigate the nature of long-lived photogenerated charge polarons ($P^{\pm}$) and neutral triplet excitons (TE) in $\pi$-conjugated polymers and oligomers [1–4]. In pristine polymers, the majority of the primary photoexcitations have been found to be singlet excitons, which may radiatively decay in the form of photoluminescence (PL) emission with high quantum efficiency. Intersystem crossing to the triplet manifold is one of the prominent non-radiative decay processes, leading to long-lived photoinduced absorption (PA) band in the near ir spectral range associated with TE. In contrast, in polymer/$C_{60}$ blends [5,6] and impurity or oxygen contaminated films [7,8], an ultrafast photoinduced charge transfer occurs that produces long-lived polarons accompanied by PL quenching. In particular, pristine derivatives of poly-p-phenylene vinylene [PPV] have efficient PL and TE; whereas the PPV derivative MEH-PPV/$C_{60}$ blends or photo-oxidized films show ***irreversible*** reduction of PL efficiency and TE, and enhanced long-lived polarons [9].

In this work we show that UV illumination (UVI) transforms a relatively pure MEH-PPV film, in which no long-lived photogenerated polarons are observed (phase *A*), into a metastable phase *B* that supports long-lived photogenerated polarons. Importantly, the films in phase *B* were found to return to their original phase A within ½ hour at room temperature in the dark; at lower temperatures this recovery is considerably longer. We advance a model in which UVI reversibly creates defects on the polymer chains that act as deep traps for the primary photogenerated polarons.

The PM spectra were obtained by measuring the differential transmission, $-\Delta T/T$ of the transmission, $T$ induced by a pump beam excitation at various temperatures, pump modulation frequencies and intensities, using a standard PM set-up [1], or a Fourier transform infrared PM spectrometer [10]. The drop cast films were made of pristine MEH-PPV with a narrow distribution of molecular weight centered at 720,000 g/mole supplied by American Dye Source (ADS).

Fig. 1 shows the 80 K PM spectrum of a MEH-PPV film kept under vacuum in the dark overnight (curve 1). The PM spectrum is typical of pristine MEH-PPV films (phase A) that is dominated by a single PA band at ~1.35 eV due to TE [9]; other PA bands due to photogenerated polarons are not detected here. Fig. 1 also contains (curve 2) the PM spectrum measured at the same experimental conditions, but after the film in phase *A* was subjected to prolonged (200 mins.) UVI of 0.1 W/cm$^2$ at 362 nm, for a total illumination dose of ~3x10$^{21}$ photons thus transforming it to phase *B*. This PM spectrum substantially differs from that of phase *A*. It shows two distinct *polaron-related spectroscopic features*, namely: (a) photoinduced ir-active vibrations (IRAV) below 0.2 eV; and (b) low-energy PA band centered at $P_1$ ~0.3-0.4 eV. In addition, the PA band related to TE is much weaker here (seen as a shoulder at ~ 1.3 eV). Also, a new photobleaching (PB) region (i.e., $\Delta T/T > 0$) appears between 0.8 to 1.1 eV that may indicate a broadening of the absorption onset in phase *B* films. The new features, namely IRAV, $P_1$ and PB scale with each other at various excitation intensities, as well as with the duration of the UVI.

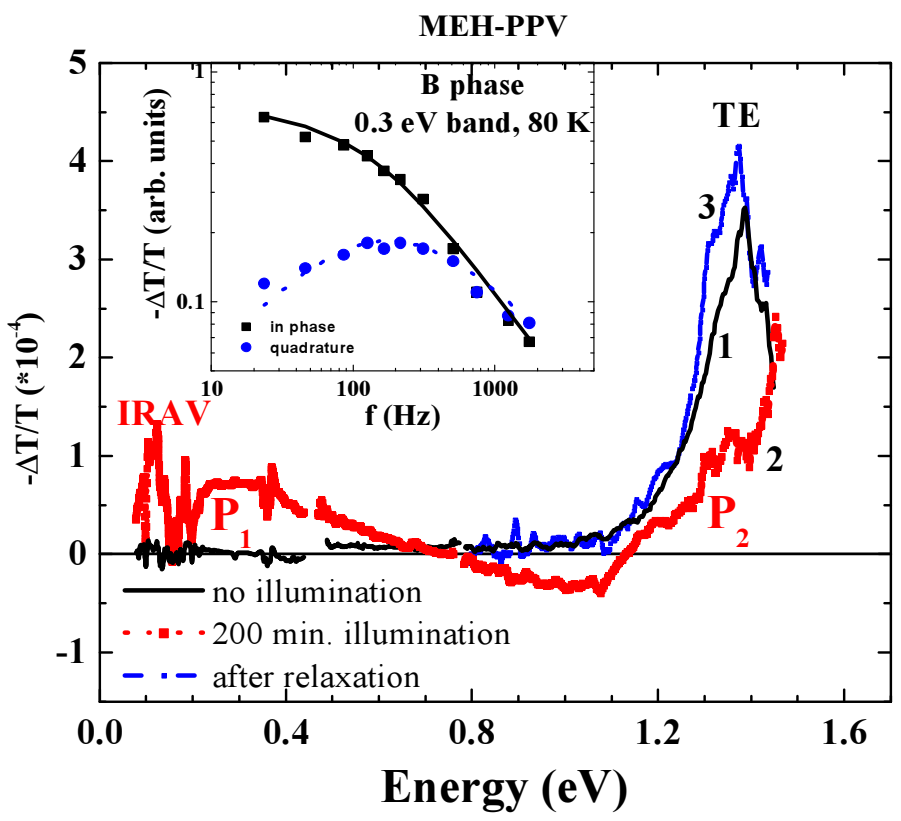

Fig. 1. PM spectra of neat MEH-PPV film at 80 K upon UVI at 360 nm and 0.1 W/cm$^2$. Curve 1 (black solid line): before UVI (phase *A*), dominated by the TE PA band. Curve 2 (red dotted line): after 200 min UVI (phase *B*), showing PA polaron features (IRAV and $P_1$ band), PB region near 1 eV, reduced TE band, and appearance of $P_2$. Curve 3 (blue dashed line): after overnight annealing at RT (phase *A* again), showing no long-lived polaron features and a recovered TE band. Inset: The modulation frequency dependence of $P_1$ in phase *B*. The solid (dotted) line is a fit of the in-phase (quadrature) component using a dispersive recombination model [12,13] with an average lifetime of 0.9 ms and dispersive exponent $\alpha \cong 0.73$.

We studied the transformation of phase *A* into phase *B* with the UVI time, t by measuring the PB and $P_1$ bands related with phase *B* at various UVI times as shown in Fig. 2 for PB(t). Fig. 2 (inset) shows the polaron induced phase *B* component, $I_B$ as measured by PB(t). It is seen that the phase transformation is slow, following a saturation-type curve $I_B \propto (1-\exp(-t/\tau_B))$, with $\tau_B$ ~ 3 hours (equivalent to irradiation of ~ $3x10^{21}$ photons at 362 nm).

The blue dashed line (curve 3) in Fig. 1 is the PM spectrum of the same film measured under the same experimental conditions, but after the film was annealed overnight, under vacuum, in the dark at room temperature. Comparing the PM spectrum of the annealed film to that of the original film, it is clear that phase *B* completely disappeared leaving the MEH-PPV film in its original phase *A* characterized by the TE PA band. We thus conclude that upon UVI the pristine MEH-PPV film in phase *A* transforms into a metastable phase *B* in a reversible manner.

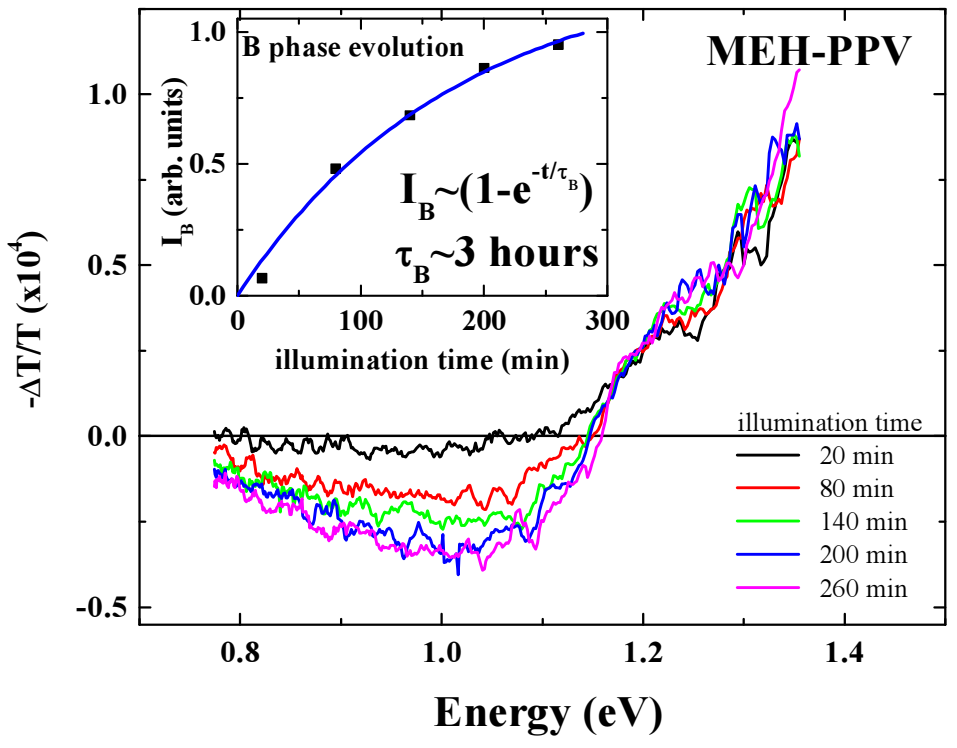

Fig. 2. The PB band related to phase *B*, at various UVI times, t as indicated. Inset: The strength, $I_B$ of the PB band vs. t. The curve through the data points is a fit to a saturation behavior (see text).

The time it takes for phase *B* films to recover to phase *A* was studied at room and liquid nitrogen temperatures. We found that ~ ½ hour at room temperature suffices to completely recover phase *A*; whereas at 80 K we could not detect any recovery even after the sample was left in the dark for over 2 hours.

As can be seen in Fig. 1, the illumination-induced phase *B* is associated with a significant reduction of the TE PA band. Since TE's are generated from singlet excitons via intersystem crossing, then the TE PA decrease in phase *B* may indicate a correlated decrease of the PL intensity. In Fig. 3 we show the effect of the UVI on the PL. It is clearly seen that the PL emission band substantially weakens upon UVI, but its spectrum does not change much. The PL decrease is summarized in Fig. 3 (inset), where the PL intensity, $I_{PL}$ is plotted vs. the UVI time, t. $I_{PL}(t)$ may be fit by the function: $I_{PL} \propto (a+\exp(-t/\tau_{PL}))$, with $\tau_{PL}$~15 mins. Comparing the rates $\tau_{PL}^{-1}$ and $\tau_P^{-1}$ (Fig. 2, inset), it appears that the PL emission decreases much faster than the appearance of the polaron PA in phase *B*. This, how-ever may be due to an increase in self absorption related to the appearance of the PB band in phase *B*, rather than to an accelerated PL decrease. Also the robust PL spectrum upon UVI shows that the PL decrease is not induced by an effect related to conjugation length interruption such as carbonyl defects; otherwise the PL spectrum would have blue shifted with UVI. Similar to the PM spectrum recovery mentioned above, the PL intensity relaxes back to the original intensity associated with phase *A* when the film is annealed in the dark at room temperature.

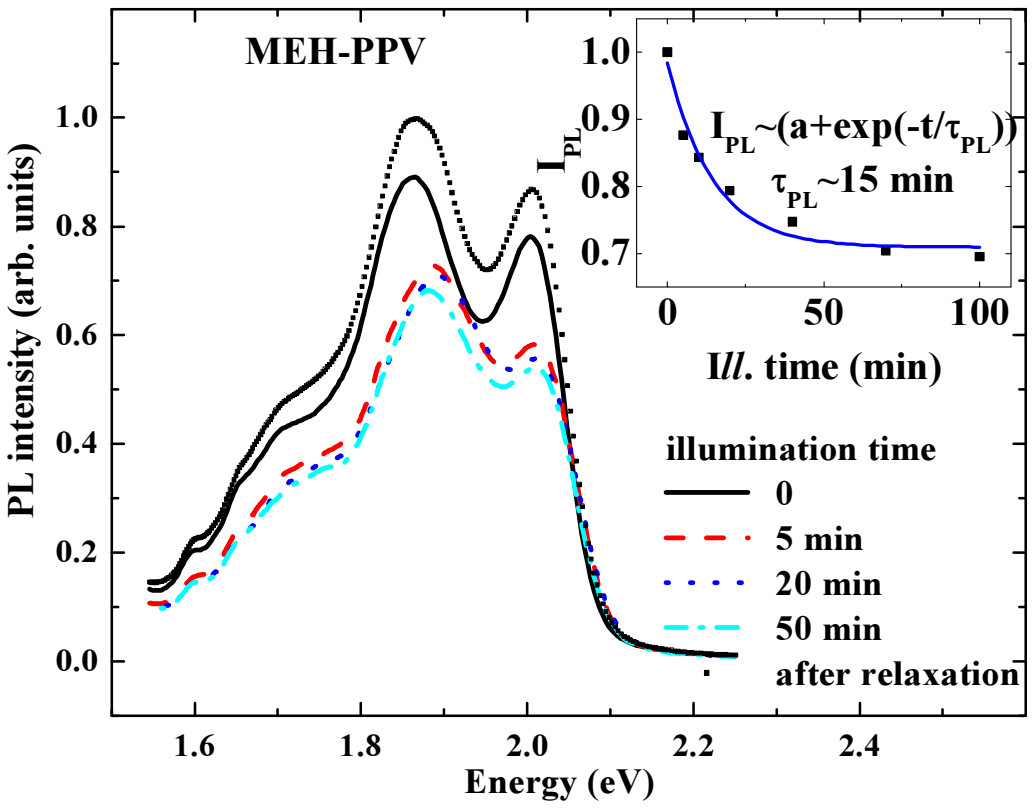

Fig. 3. PL spectra of MEH-PPV film at 80 K at 360 nm and 0.1 W/cm$^2$ for UVI times, t as indicated. Inset: The PL intensity, $I_{PL}$ vs. t; the curve through the data points is a fit to $I_{PL}$ (see text).

The steady state PA signal is proportional to the product of the photogeneration quantum efficiency (QE), and the lifetime ($\tau$) of the relevant photoexcited species:

$$PIA \propto I_L \cdot QE \cdot \tau \ , \qquad (1)$$

where $I_L$ is the pump intensity. Thus, the long-lived polaron PA bands seen in phase *B* may originate from either a large increase in QE, or a dramatic increase in the lifetime of short lived polarons that are originally photogenerated in phase *A*. In order to separate these two possible underlying causes, we measured the polaron dynamics in phase *B* and compared it with the polaron dynamics of polarons in phase *A* [11]. Fig. 1 (inset) shows the modulation frequency dependence of the lower polaron band ($P_1$) in phase *B* measured at 0.3 eV. The average lifetime, $\tau_0$ is extracted directly from the frequency $f_{max}$ (~200 Hz) at the maximum of the quadrature component; we get $\tau_0 \sim 1/(2\pi f_{max}) \sim 1$ ms. Noting also that the $P_1$ frequency dependence is strongly indicative of dispersive recombination [12,13], we show in Fig. 1 (inset) fits using the dispersive recombination model. These fits result in a similar $\tau_0$. On the other hand, previous time resolved PM measurements of pristine MEH-PPV in phase *A* revealed the existence of short-lived polarons with lifetime ~ 1 ns [11]. We thus conclude that the polaron lifetime in phase *B* films is about $10^6$ times longer than that of free polarons in phase *A* films. This strongly indicates that the enhanced polaron PA in the PM spectrum of phase *B* is caused by *lifetime increase* rather than an induced increase in photogenerated polaron quantum efficiency. The dramatic polaron lifetime increase in phase *B* is probably caused by charge trapping into efficient deep traps that are formed in the polymer chains upon UVI. The following model may cast light on the origin of these deep traps.

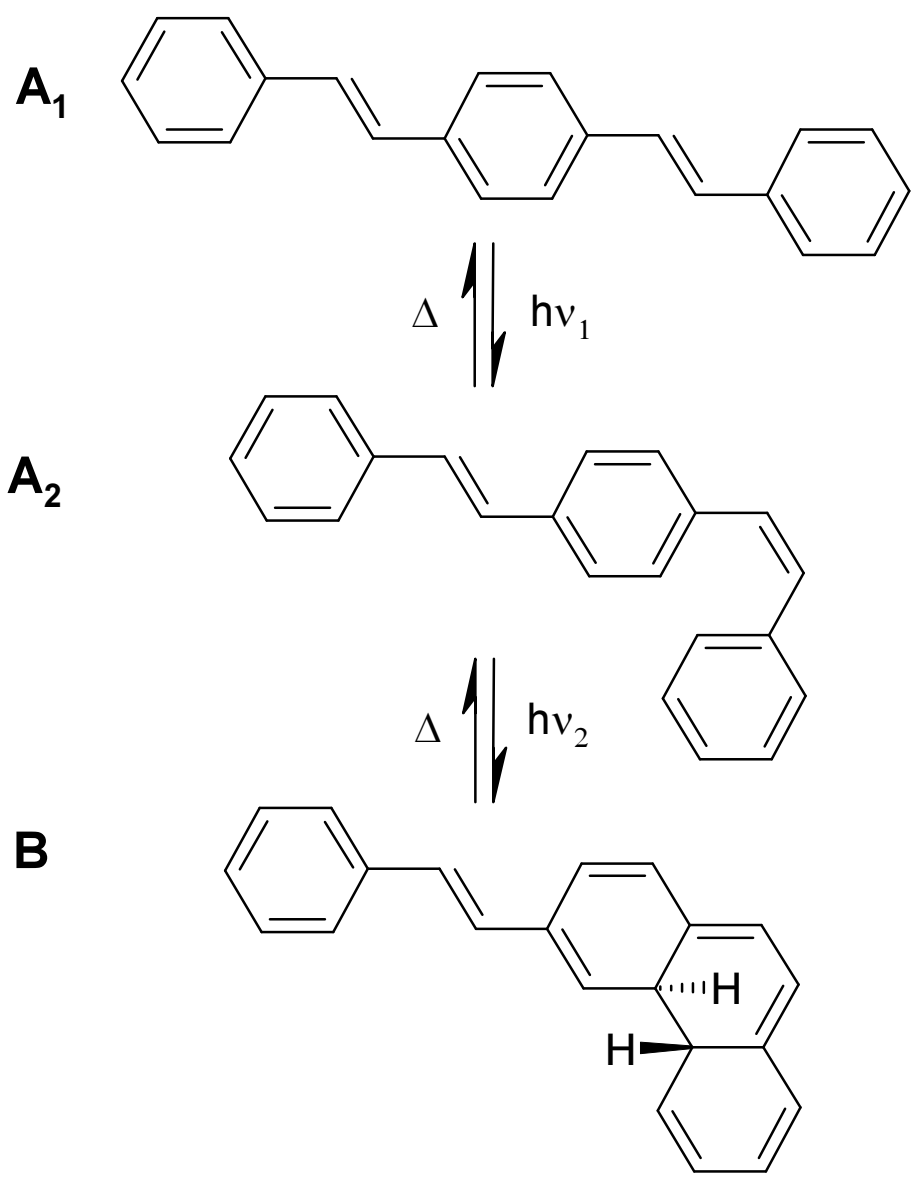

Fig. 4. A possible reversible mechanism for the UVI-induced "polaron-supporting phase" [phase *B*]. Shown schematically are the relevant segments of MEH-PPV. $A_1$: all-trans, $A_2$: native cis-defect on the trans chain. Both $A_1$ and $A_2$ constitute the pristine *A* phase. B: UVI- induced charge trap that dominates phase *B* films.

Fig. 4 depicts a possible mechanism for the UVI-induced phase *B*. In analogy to the well-documented photocyclization of stilbene deriva-tives into dihydrophenanthrene derivatives [14], we propose that Z-diphenyl ethene (cisdiphenyl ethene) conformer regions of the polymer chain are pre-organized to undergo a photochemically allowed cyclization into dihydro phenanthrene defects. These types of metastable species are normally characterized by lifetimes that may vary from seconds to months at room temperature. In the dark, the dihydrophenanthrene skeleton undergoes a thermally induced ring opening, thus reverting to the Z-diphenyl ethane structure. An intensively studied process is the photocyclization of stilbene derivatives into dihydrophenanthrene derivatives. In this process, upon photon absorption an E-diaryl ethene (trans-diary ethene), Fig. 4($A_1$), transforms into its Z-conformer, Fig. 4($A_2$). This process may be slowly reverted thermally or photochemically, by photon absorbing into the Z-diarylethane conformer (Fig. 4($A_2$)). Pristine MEH-PPV in "Phase A" constitutes mainly of $A_1$ species, but may also include segments of $A_2$ species. The photoexcited Z-diarylethane conformer (Fig. 4($A_2$)) may also undergo a photochemically allowed photocyclization to form the dihydrophenanthrene photoproduct, Fig. 4(B). The latter species is metastable, reverting thermally to the Z-diarylethene conformer with a characteristic lifetime that may vary from seconds to months, and even years at room temperature, depending on the exact molecular structure [14]. MEH-PPV in Phase B above includes segments of B species due to photocyclization. This kind of **reversible** photocyclization was reported to occur in oligomers as well as in polymeric species [14,15]. The dihydrophenanthrene photoproduct (Fig. 4(B)) acts as a trap for either positive or negative charge carriers, leaving behind long lived opposite-charge polarons on the MEH-PPV polymer chains.

In summary, we discovered a new UVI induced metastable polaron supporting phase in neat MEH-PPV films. The pristine films in their original phase *A* show strong PL emission and long-lived photogenerated triplet excitons, but no long-lived polarons. Prolong exposure to UVI reversibly transforms the pristine films into a different phase *B*, in which long-lived polarons are photogenerated and the PL emission is considerably reduced. The phase *B* films transform back to their original phase A with a temperature dependent recovery time of ~ ½ hour at room temperature, and many hours at liquid nitrogen temperature. We propose a plausible mechanism by which UVI reversibly transforms native defects in the polymer chains into deep charge-traps.

**Acknowledgments**–Work supported by the Israel Science Foundation 735/04, and DOE FG-02-04 ER46109. ZVV acknowledges support from the Lady Davis foundation at the Technion.

---